# Information Behavior During the Covid-19 Crisis in German-Speaking Countries

Stefan Dreisiebner[a,b]*, Sophie März[a] and Thomas Mandl[a]

[a] *Department for Information Science and Natural Language Processing, University of Hildesheim, Hildesheim, Germany*

[b] *Department of Corporate Leadership and Entrepreneurship, University of Graz, Graz, Austria*

\* Elisabethstraße 50b, 8010 Graz, Ausria; stefan.dreisiebner@uni-graz.at

# Information Behavior During the Covid-19 Crisis in German-Speaking Countries

**Abstract**

**Purpose** – Exploring the impact of the Covid-19 crisis at the level of individual information behavior among citizens from the German-speaking countries Austria, Germany and Switzerland.

**Design/methodology/approach** – An online survey was conducted among 308 participants gathered through convenience sampling in April and May 2020, focusing on how citizens changed their mix and usage intensity of information sources and according to which criteria they chose them during the Covid-19 crisis. A Wilcoxon signed rank test was used for testing central tendencies. Effect sizes were considered to support the interpretation.

**Findings** – The results show first that the Covid-19 crisis has led to an increased demand for reliable information. This goes alongside a significant increased use of public broadcasting, newspapers and information provided by public organizations. Second, the majority (84%) of the participants reported being satisfied with the information supply during the Covid-19 crisis. Participants who were less satisfied with the information supply used reliable sources significantly less frequently, specifically public television, national newspapers and information provided by public organizations. Third, the amount of Covid-19-related information led some participants to a feeling of information overload, which resulted in a reduction of information seeking and media use.

**Originality** – This paper is one of the first to analyze changes of information behavior patterns of individuals during crises in the current information environment, considering the diversity of resources used by individuals.

**Introduction**

The disease Covid-19, caused by the beta coronavirus type SARS-CoV-2, was first described at the end of 2019 in Wuhan, in the People's Republic of China (PRC), and by January 2020 had developed into an epidemic throughout the country. The first confirmed Covid-19 case in the German-speaking countries was registered by January 27, 2020 in Germany, and then on February 25, 2020 in both Austria and Switzerland. On March 11, 2020, Covid-19 was declared a global pandemic by the World Health Organization (WHO). To slow the spread of the virus, far-reaching movement restrictions for all citizens were announced by the federal governments in Austria, Switzerland and Germany on March, 13,16 and 22, 2020, respectively. Where possible, employees were recommended to work from home. Special precaution was recommended for people over 65 years and people with specific existing health conditions, who were considered as high-risk groups for Covid-19 (Robert Koch Institute, 2020). The Covid-19 crisis has led to a great demand for information. To date, large-scale patterns have been observed, which show that global Internet traffic has increased as well as the number of visits to news sites, in particular (Siddique *et al.*, 2020). *The New York Times* reported that in the USA, established national and regional newspapers received more than twice as many visits in March than in February. In the same period, partisan and polarizing sites received fewer visits (Koeze and Popper, 2020).

For the situation in the German-speaking countries, an increase in Alexa rankings of several sites was observed. Regional public news channels greatly gained in popularity between January and March as well as the website of the national public radio (DLF). On the other hand, the Alexa popularity rankings dropped in the same time period for sites with less recent information (e.g. Wikipedia) and even for general search sites (e.g. Google). Those users searching on Google posed more queries on sites with reliable information (e.g. main evening news programs, the Robert Koch Institute, RKI), as a look at Google Trends shows. The WHO has warned that much health misinformation is being distributed concerning Covid-19, and has referred to the pandemic as also being an *infodemic* (Richtel, 2020).

In this study, we investigate individual information behavior during the crisis. There have been studies on information behavior and information dissemination for other crises. However, the information ecology is constantly changing, and it seems necessary to analyze the effect of the pandemic on the current information environment. The dissemination of information in social media, and in particular on Twitter, has already been researched in the first months of the crisis. However, the focus of such studies is mainly on big data and does not reflect the diversity of resources used by individuals. We intend to show how citizens change their mix and usage intensity of resources and according to which criteria they are choosing their information sources. This leads us to the following research questions: (1) Have citizens in the German-speaking countries changed their mix and usage intensity of information sources in the Covid-19 crisis? (2) According to which criteria are they choosing their information sources during the Covid-19 crisis?

**Literature Review**

Information behavior refers to 'those activities a person may engage in when identifying his or her own needs for information, searching for such information in any way, and using or transferring that information' (Wilson, 1999, p. 249). When searching for information is conducted purposefully with the objective of finding information about health, it can be defined as health information seeking behavior (Zimmerman and Shaw, 2020).

*Media Research in the German-speaking Countries*

Information behavior is related to media research, which focuses on 'any form of communication transmitted through a medium (channel) that simultaneously reaches a large number of people' (Wimmer and Dominick, 2013, p. 2). There is a vast amount of research in this area, including from the private sector, such as in-house research conducted by media companies (Wimmer and Dominick, 2013). This review focuses on studies from the German-speaking countries. An extensive longitudinal study is the ARD/ZDF study on mass communication, conducted by the German public broadcasters ARD and ZDF since 1964. The most recent study, conducted in 2019, found that TV and radio stations are facing increasing competition from video streaming services. Germans between 14 and 29 years old are already predominantly using streaming services, while for the overall German population regular TV and radio still dominate. On average, Germans spend seven hours a day consuming various media, while the group below 30 years only consumes six hours a day (Frees *et al.*, 2019). The study further found that Germans especially value public broadcasters for their journalistic quality, regional news coverage and their support for political opinion-building, with those under 30 years of age in particular valuing the services of public broadcasters positively. In contrast, privately owned TV and radio stations are mainly consumed for the purpose of entertainment (Breunig and Hottmannspötter, 2019). A study by PricewaterhouseCoopers (PwC, 2018) among Germans (N=1,000) showed that only 18% trust information published on Facebook and 85% would wish for more staff and resources to allow media a more high-quality journalistic coverage. In the German-speaking countries, the legally required objective of public broadcasting is to make high-quality programs, supply good information and involve people in a democratic culture, while privately owned broadcasters focus on commercially attractive audience groups (Meijer, 2005).

### *Information Behavior During (health-related) Crises*

The information needs of humans change during existential crises (Butenaitė *et al.*, 2016). Basic needs like safety and assurance of survival become relevant and will change the content the humans seek. When basic needs are fulfilled, citizens want to achieve other goals in their lives, but in a dangerous situation with many deaths, the information needs focus on own safety, support for others and often the expression of emotions. The changing information ecology requires that this emergency behavior is studied in the context of recent crises.

Liu *et al.* (2016) predicted social media use during disasters, based on a field experiment simulating a hypothetical disaster. They found that upon initial exposure to information about the disaster, the participants reported intentions to communicate about it predominantly via interpersonal forms, such as telephone calls, face-to-face conversations, texting and emailing, rather than through social media. They also found that participants were significantly more likely to seek further disaster information via TV after the information initially came from a national government web post. van der Meer and Jin (2020) conducted an online experiment to seek a formula for misinformation treatment in a public health crisis. They found that the presence of corrective information leads to the debunking of incorrect beliefs based on misinformation, and that government agency and news media sources are more successful in improving belief accuracy compared to social peers. Similarly, Southwell

*et al.* (2016) found based on analysis of relationships between news coverage, social media mentions and online search behavior regarding the Zika virus in the USA, Guatemala and Brazil, that news coverage of public health authority announcements leads to information sharing, engagement and searching. Their results also showed strong positive correlations between news (daily volume) and Twitter posts as well Google searches for all three countries.

      Sommerfeldt (2015) examined how Haitians used information source repertoires to meet information insufficiencies following the 2010 earthquake based on surveys. He found two different repertoires of media: on the one hand a traditional repertoire of radio, TV, church and word of mouth; and on the other, an elite repertoire of newspapers, the Internet, short message service (SMS), billboards and the national police. People with higher education were more likely to use the elite repertoires. Gender differences were also observed, with men being more active in seeking. Focusing more on the technical perspective regarding the problem of identification of fact-checkable tweets posted during a particular disaster event, Barnwal *et al.* (2019) described the findings of focus groups using Twitter data on the 2015 Nepal earthquake, leading to an algorithm that provides highly fact-checkable tweets containing specific references.

      Jang and Baek (2019) studied how the credibility of information from public health officials is associated with people's reliance on a particular communication channel in the context of the 2015 Middle East Respiratory Syndrome (MERS) outbreak in South Korea. They found that less credible information from public health officials led to more frequent use of online news, interpersonal networks and social media for acquiring MERS-related information. However, they found no relation between the

credibility of information from public health authorities and the use of television news or print newspapers.

*Information Behavior During the Covid-19 Crisis*

For the current Covid-19 crisis, there have already been several data collections curated (e.g. Chen *et al.*, 2020; GESIS, 2020; Lamsal, 2020) and some first results of data-oriented studies have been published (for a first overview see Siddique *et al.*, 2020). Information propagation on social media has revealed several interesting preliminary results. A substantial amount of potentially incorrect information is being disseminated during the Covid-19 crisis (Yang *et al.*, 2020). Also, artificially created information by bots is having an impact on the spread of misinformation (Ferrara, 2020). The amount of ethnic hate and aggressiveness has also increased, according to an analysis by Stechemesser *et al.* (2020).

Ali *et al.* (2020) reports of an online survey among 11,242 US adults, that was conducted in two rounds during March and April 2020 using advertisement-based recruitment on social media. The results show that traditional media sources (television, radio, podcasts, or newspapers) were the largest sources of Covid-19 information (91.2%) and that the largest individual information source – and also most trusted one – was government websites (87.6%). Additionally, they found that participants who were male, aged 40-59 years or ≥60 years; not working, unemployed, or retired; or Republican were likely to use fewer sources while those with children and higher educational were likely to use more sources. They also found significant associations between information source and Covid-19 knowledge. Soroya *et al.* (2021) examine the factors leading to information avoidance during Covid-19 pandemic, based on 321 responses gathered from students, staff, and faculty members of three Finnish universities during April 2020. They found that mass media, print media and official websites were used for information seeking during the Covid-19 pandemic, but only social media exposure resulted in information overload and information anxiety.

First studies have already been published regarding the situation in the German-speaking countries. Schäfer *et al.* (2020) analyzed the health information-seeking behavior of 443 German university students, based on two cross-sectional surveys in the summer of 2019 and 2020. The results show that students' health information seeking takes place primarily online and changed remarkably during the Covid-19 crisis: use on online radio almost tripled and use of online news sites and social media almost doubled compared to before the crisis, while the use of search engines, encyclopedias, health portals and similar websites with less recent information dropped. Additionally, significant correlations between the intensity of Covid-19-related information seeking, risk perception, and actual risk behavior were found. The authors raise their concern wether the comparatively high relevance of sources that are largely based on unchecked user-generated guarantees the necessary quality and reliability of health information. Dadaczynski *et al.* (2021) focused on web-based information seeking behavior, based on a web-based survey among 14,916 university students aged over 18 years from 130 universities across all 16 federal states of Germany. The findings show, that search engines, news portals, and websites of public bodies were most often used as information sources on Covid-19. The study also assessed the health literacy of participants, which showed the greatest difficulties for assessing the reliability of health-related information and the ability to determine whether the information was written with a commercial interest. A low ability to critically evaluate information did go along with use of social media, while the opposite was observed for the use of public websites.

However, these previous studies in the German-speaking countries did only focus on students and online information sources. There is a need to analyze the perspective of individuals on a broader basis to see how citizens change their mix and usage intensity of resources and according to which criteria they are choosing their information sources.

**Methodology**

Data on opinions and perspectives of individuals regarding their information behavior during the Covid-19 crisis were gathered through an online questionnaire (Raab-Steiner and Benesch, 2018). An online questionnaire was used due to the ease of distribution and the possibility of reaching out to all German-speaking countries. It was expected that during the lockdown people would be more open towards filling out surveys. The questionnaire was implemented in the German language in LimeSurvey, hosted by the University of Hildesheim.

The questionnaire consisted of several parts: At the beginning, the participants were introduced to the aims of the study and asked to agree to the data privacy statement. The data privacy statement informed participants that anonymity is assured, how the collected data is being stored, who has access to the data, precautions for data protection and the participants' rights regarding the gathered data in accordance to the European Data Protection Regulation. This was followed by questions on demographic data, such as age and gender, whether the participants currently worked from home and whether they thought they were part of a high-risk group for Covid-19. After that, they were first asked whether they personally felt they were consuming more news during the Covid-19 crisis. Second, they were asked how intensively they used various sources of information (as listed in Table II within the results-section) at the time of participation (during the Covid-19 crisis), how much they used the same sources before the crisis and whether they were using any specific source that they had not consulted before the crisis. Third, they were questioned on the basis of which specifically listed criteria they were choosing their sources of information in the Covid-19 crisis. Fourth, they were asked how satisfied they were with the supply of information about the crisis and what were the main reasons for looking for information during the crisis. Fifth, there was the opportunity to leave a general comment if they had noticed any other changes in their media use and information and communication behavior since the beginning of the crisis and there was the opportunity to add any further comments or suggestions they might have regarding the study. Only the questions of the information behavior section of the questionnaire were mandatory. According to German law this non-interventional survey design did not require prior approval by the university ethics committee. The regulations of the University of Hildesheim required the study to be approved by the data protection officer. This approval was given.

The items were designed based on an extensive literature review. The aim was to use existing items that have been already proven in terms of reliability. The items on media use were adapted from Liu *et al.* (2016) and PwC (2018), and the criteria for media usage adapted from Breunig and Hottmannspötter (2019).

Most questions allowed answers on a 7-point Likert scale, asking either for the frequency (1: Daily; 2: Two or more times per week; 3: Once per week; 4: Two or more times per month; 5: Once per month; 6: Less than once per month; 7: Never) or the agreement (1: Applies to a great extent; 2; 3; 4: Neutral; 5; 6; 7: Does not apply at all). There were also some open questions, allowing the participants to justify their answer or to provide additional remarks.

Pretests were conducted to ensure a proper understandability, usability and length of the online questionnaire. Three people were asked to complete the draft of the questionnaire. Afterwards they were interviewed by phone. The participants were chosen to represent different levels of age and education: a 26-year-old female (university student); a 29-year-old male (university graduate working in the IT sector); and a 54-year-old female (with no college education, working in consulting). The findings from the pretest led to the reformulation of some items to enhance understandability, e.g. the addition of examples for some information sources.

The sample was obtained through convenience sampling (Galloway, 2005) using different distribution channels, in order to reach different cohorts of media users and educational levels among the adult population of the German-speaking countries: The questionnaire was initially promoted and distributed through mailing lists within the German-speaking information science community and on social media (Facebook, Twitter, LinkedIn), where the authors initially shared the survey with their personal profiles, and then it was later forwarded and retweeted by other users. To also reach traditional media consumers, additional support in the promotion was given by the media office of the University of Hildesheim, which resulted in reports on the project and invitations to participate being posted on the university website and in local and national media. The survey was conducted from April 12, 2020 until May 17, 2020.

The acquired data were afterwards analyzed through Microsoft Excel and IBM SPSS Statistics. As suggested by Sullivan and Artino (2013), means were computed for the Likert-type scales. A Wilcoxon signed rank test was used for testing central tendencies. Effect sizes (Cohen, 1988) were considered to support the interpretation.

**Results**

In total, 308 people participated in the online survey. Table I gives an overview of the demographic data of the participants. Comparing the composition of the sample of participants reporting working from home and viewing themselves as part of a high-risk group for Covid-19 to the full sample shows only some noteworthy differences. Participants reporting working from home were older, with more participants from the group between 50 and 59 years of age (31%) and over 60 years of age (33%). They also worked less from home during the crisis (36% vs. 22%), which might be due to the higher age and thus a higher number of retirees in the sample.

[Table I near here]

The demographic distribution of our sample is not representative for the German-speaking countries: According to data from the national statistical bureaus, in Austria, Germany and Switzerland female citizens account for 51% and male citizens for 49% of the society. Thus, the share of female participants is higher in this sample. Regarding the age, the group <18 ($M$ for the population of the German-speaking countries = 18%) and >60 ($M$ = 26%) are underrepresented, while the groups 18-29 ($M$ = 13%), 30-39 ($M$ = 14%), 40-49 ($M$ = 13%) and 50-59 ($M$ = 16%) are overrepresented. Geographically, the federal state of Lower Saxony is overrepresented, which only accounts for 9.6% of German citizens. The federal states of Bavaria and Berlin are relatively well represented, while the other regions are underrepresented. This might be due that most media promotion for the study was achieved by media based in Lower Saxony. Finally, participants in this sample are higher educated than the average society. While within the German-speaking countries the median share of citizens with university degree is 20%, 70% of the participants claimed to have graduated university. Lower educational levels are underrepresented.

*Media Usage Behavior Regularly & During the Crisis*

Of the total number of participants, 75% felt that overall they had consumed more news and information since the beginning of the Covid-19 crisis. Table II provides an overview of how the participants felt that their media use changed on a more detailed basis. As can be seen, the most used channels during the crisis were online communication with acquaintances and friends ($M = 2.16$, $SD = 1.45$), followed by public television ($M = 2.24$, $SD = 1.79$), national newspapers (offline & online) ($M = 2.51$, $SD = 1.90$) and local newspapers (offline & online) ($M = 3.06$, $SD = 2.14$). In contrast, the least used channels during the crisis were private television ($M = 5.13$, $SD = 2.23$) and the social media channels Instagram ($M = 5.08$, $SD = 2.57$), Twitter ($M = 4.93$, $SD = 2.66$) and Facebook ($M = 4.85$, $SD = 2.52$).

    A Wilcoxon signed rank test shows that the only channels that decreased highly significantly were physical meetings and discussions ($z = -11.47$, $p < .001$). The effect size (Cohen, 1988) was $r = .65$, which is a strong effect. No change is visible for the social media channels YouTube, Facebook and Instagram. All other channels show at least a significant increase. The highest increase ($MDiff = 2.54$), which is highly significant ($z = -11.47$, $p < .001$), is visible for public organizations (like the Robert Koch Institute and the Federal Office of Public Health). The effect size was $r = .77$, which is a strong effect. A highly significant increase with a medium effect is also shown for public television ($z = -6.51$, $p < .001$, $r = .37$), international sources (radio, broadcast, newspaper) ($z = -6.16$, $p < .001$, $r = .35$), national newspapers (offline & online) ($z = -5.95$, $p < .001$, $r = .34$), online communication with acquaintances and friends ($z = -5.62$, $p < .001$, $r = .32$) and local newspapers (offline & online) ($z = -4.38$, $p < .001$, $r = .25$). A significant increase with a small effect is shown for private television ($z = -2.99$, $p = .003$, $r = .17$), Twitter ($z = -2.89$, $p = .004$, $r = .16$) and music streaming and podcasts ($z = -2.41$, $p = .016$, $r = .14$). Thus, Twitter is the only social media channel that saw a significant increase in use among the sample.

    [Table II near here]

Figure 1 shows the distribution in percent for each of the items of the Likert scale for the media channels with considerable changes during the Covid-19 crisis. As can be seen, changes happened mainly between irregular users, who changed to daily users, while non-users continued to stay away. The only deviation from this behavior pattern is seen for public organizations, where the participants declaring to never use them decreased from 44% to 8%, with almost 20% claiming to visit them daily during the crisis.

[Figure 1 near here]

This part of the questionnaire closed with an optional open question: 'What specific source are you using now that you did not consult before the crisis?' An answer was provided by 255 participants. These answers are in accordance with the previously reported development in media use: The most mentioned source was the German Robert Koch Institute (92 mentions). The second most mentioned source was podcasts (51 mentions), especially those provided by the German public broadcasters NDR and RBB, and also featuring the German virologist Christian Drosten. He was the third most mentioned source (44 mentions). Fourteen participants reported using YouTube, especially to watch the YouTube channel of the German public television ARD and interviews with experts. Ten persons mentioned Twitter as an additional source, which was described as being used to follow accounts of potential information sources, such as virologists, medical journalists, local newspapers, the Robert Koch Institute, the Federal Statistical Office or the Federal Office of Public Health.

*Criteria for Choosing Information Sources*

The participants were overall satisfied with the information supply during the crisis. Eighty-four percent reported that they were satisfied, 4% were neutral and 12% were unsatisfied. Figure 2 shows the reported criteria that were personally important when choosing sources of information during the Covid-19 crisis. The most important factor was credible information ($M = 1.16$, $SD = .596$), followed by journalistic quality ($M = 1.39$, $SD = .84$), interesting facts from research, history ($M = 1.81$, $SD = 1.02$), information from official sources ($M = 1.9$, $SD = 1.29$), background information on many topics ($M = 2.28$, $SD = 1.2$) and availability of topics from one's own region ($M = 2.44$, $SD = 1.39$). The least important factor was knowing the author personally ($M = 5.45$, $SD = 1.68$), followed by good entertainment ($M = 4.58$, $SD = 1.66$), personal recommendation ($M = 4.31$, $SD = 1.77$) and interesting for the whole family ($M = 4.26$, $SD = 1.87$).

[Figure 2 near here]

*Reasons for Information Seeking During the Covid-19 Crisis*

Figure 3 gives an overview of the reasons for information seeking, as reported by the participants. The most common motivation was to monitor the general situation, which was stated by 96% of the participants. This was followed by gathering information on economic and social aspects of the crisis (80%), obtaining information on movement and travel restrictions (70%) and avoiding infection (59%). Only 4% of the participants reported having actively avoided information and reports on Covid-19. Twenty-four participants also stated other reasons through a text field. These answers included following world affairs, gathering knowledge to support their own opinion-building, concern for family, behavior suggestions and impact of the crisis on the personal level as well as for society.

It also shows that, for participants counting themselves as being in a high-risk group, personal health concerns (77%) and avoiding infection (79%) were considerably more important, while information on economic and societal aspects of the crisis was of less importance (72%). None of the participants counting themselves as being in a high-risk group reported having actively avoided information and reports on Covid-19.

[Figure 3 near here]

*Further Changes in Media Use and Information and Communication Behavior*

The questionnaire finished with an open question, providing the participants the opportunity to describe any further changes in media use and their information and communication behavior they might have noticed during the crisis. This opportunity was used by 63% (n = 193) of the participants. Several participants reported a higher frequency in media usage and information seeking. Most commonly described were different variations of information overload. Participants reported that they were overwhelmed with the amount of information, which led to a reduction in information seeking after a phase of intense media use. Participants also described a feeling of frustration through information overload. Some of them also reported avoiding specific information sources, such as extended special editions of the evening news or live news tickers with the newest developments, or a limited media use at specific time slots. Other participants described information overload through a rise in social media use by their friends, including the forwarding of videos and social media challenges, which resulted in participants ignoring their messages. Some participants reported that they consumed other media channels like newspapers solely by following their social media appearance.

Participants also commonly reported that they reflect on information more critically since the beginning of the Covid-19 crisis. Participants stated that they actively compare information from different sources, prefer reliable information sources and are aware of the circulation of Covid-19-related misinformation. Some participants also described actively using foreign media to gain a broader view or actively reducing social media channels where they perceive the information to be less reliable.

Participants also reported different changes in the intensity of their communication – while some reported being much more in touch with their family and friends due to increased spare time and video conferencing, others described having far fewer social interactions.

### *Correlations of Reported Media Use and Age, Education and Usage Criteria*

Table III gives an overview of correlations between the reported media use during the Covid-19 crisis and criteria for choosing information sources with the age, education and satisfaction with the information supply during the Covid-19 crisis. The results show that older participants significantly more frequently used local newspapers ($r = -.187$, $p = .001$), public television ($r = -.170$, $p = .003$) and national newspapers ($r = -.112$, $p = .05$) during the Covid-19 crisis. On the other hand, Instagram ($r = .324$, $p < .001$), music streaming, podcasts ($r = .247$, $p < .001$), YouTube ($r = .197$, $p = .001$) and Facebook ($r = .160$, $p = .005$) were used significantly less frequently by older participants during the Covid-19 crisis. Between education and media use there is only one significant correlation: Higher educated participants used less frequently online communication with acquaintances and friends ($r = .113$, $p = .048$). Significant correlations appear regarding the satisfaction with the information supply during the Covid-19 crisis. Participants who were less satisfied with the information supply during the Covid-19 crisis used significantly less frequently public television ($r = .228$, $p < .001$), national newspapers ($r = .208$, $p < .001$), information provided by public organizations ($r = .169$, $p = .003$) and Twitter ($r = .157$, $p = .006$).

The results show further that older participants considered it significantly more important to know the author personally ($r = -.156$, $p = .006$), that the information source is independent from state, politics and economy ($r = -.144$, $p = .012$) and provides background information on many topics ($r = -.123$, $p = .031$). On the other hand, older participants considered it significantly less important to have current live information ($r = .197$, $p = .001$) and information from official sources ($r = .161$, $p = .005$). Higher educated participants considered it significantly less important that the content is interesting for the whole family ($r = .234$, $p < .001$), independent from state, politics and economy ($r = .173$, $p = .002$), knowing the author personally ($r = .158$, $p = .005$), personal recommendation of information sources ($r = .128$, $p = .025$) and low degree of advertisement ($r = .121$, $p = .043$).

Participants who were less satisfied with the information supply during the Covid-19 crisis considered it significantly more important that all opinions can be expressed ($r = -.263$, $p < .001$), that the information source is independent from state, politics and economy ($r = -.223$, $p < .001$), to know the author personally ($r = -.160$, $p = .005$), that the content is interesting for the whole family ($r = -.121$, $p = .033$) and to have covered topics from the own region of residence ($r = -.120$, $p = .036$). On the other hand, participants who were less satisfied with the information supply during the Covid-19 crisis considered it significantly less important that information is provided from official sources ($r = .178$, $p = .002$).

[Table III near here]

**Discussion**

The results confirm the initially described observation from global Internet traffic, that the Covid-19 crisis has led to an increased demand for information. Seventy-five percent of the participants in the survey described that, overall, they are consuming more news and information since the beginning of the Covid-19 crisis.

As initially observed through the Alexa rankings for the German-speaking countries and Google Trends, the demand increased for sources with reliable information. Participants reported having used public organizations (like the Robert Koch Institute and the Federal Office of Public Health), public television, international sources (radio, broadcasting, newspapers), national newspapers and local newspapers highly significantly more than before the crisis. This change happened – except for public organizations – mainly among irregular users, who changed to daily users, while non-users continued to stay away. This shows that citizens seem to rely on familiar sources when it comes to health crisis situations. This might be also connected with the demand for reliable information, which was also shown by the criteria the participants reported as being personally important when choosing sources of information during the Covid-19 crisis. The most important criteria when choosing sources of information during this period were credible information, followed by journalistic quality, interesting facts from research, and information from official sources. Previous research has shown that Germans especially value public broadcasters for their journalistic quality (Breunig and Hottmannspötter, 2019). As in the longitudinal study on mass communication conducted by the German public broadcasters ARD and ZDF (Frees *et al.*, 2019), the data in this study also showed that older participants used newspaper and television more frequently, but streaming services and social media less so. The high relevance of public organizations as information source during Covid-19 in Germany was also already reported by Dadaczynski *et al.* (2021). Additionally, an analysis of media coverage von Covid-19 in Swiss media showed that public broadcasting provided the highest quality on Covid-19 coverage among all media outlets, specifically in terms of variety of topics and their relevance, presence of experts, as well es contextualization (Eisenegger *et al.*, 2020).

The results show that the most common motivation for information seeking was to monitor the general situation, followed by gathering information on economic and social aspects of the crisis, obtaining information on movement and travel restrictions and how to avoid infection. Only 4% of the participants reported having actively avoided information and reports on Covid-19. For participants counting themselves as being in a high-risk group for Covid-19, personal health concerns and avoiding infection were considerably more important, while information on economic and societal aspects of the crisis were less so. Already previous research on health information seeking behavior showed that awareness of health risks is associated with perceived importance of health-related knowledge acquisition (Hay *et al.*, 2012; Rimal, 2001). Some participants reported a feeling of information overload, which resulted in a reduction of information seeking and media use or limiting the use to specific times of the day only. Such information avoidance as a result of information overload was previously described by Lee *et al.* (2017) and for the Covid-19 pandemic by Soroya *et al.* (2021).

Comparing the results of this study to previous findings from research on disasters and public health emergencies, it shows that similarly to the Zika virus outbreak in 2016 (Southwell *et al.*, 2016), information seeking grew when the Covid-19 pandemic began reaching the German-speaking countries and was thus increasingly covered by news outlets. However, in contrast to the field experiment by Liu *et al.* (2016) simulating a hypothetical disaster, participants did not react by predominantly communicating via interpersonal forms, but instead turned primarily to traditional, reliable public broadcasters and information sources. This differs from the health information seeking behavior in everyday situations, where the Internet plays an important role and online sources are trusted as being reliable (Beck *et al.*, 2014; Aoun

*et al.*, 2020). However, the Covid-19 pandemic cannot be completely compared to an unexpected disaster where interpersonal communication and social media might be the only available information sources in the immediate time after the event. In contrast to Jang and Baek (2019), who found that people in South Korea used online news, interpersonal networks and social media after receiving less credible information from public health authorities during the MERS outbreak, such an effect could not be seen within this sample. Social networks were reported to be among the least important information sources during the Covid-19 crisis.

However, the vast majority (84%) of the participants reported being satisfied with the information supply during the Covid-19 crisis. Participants who were less satisfied with the information supply during the Covid-19 crisis used reliable sources significantly less frequently, specifically public television, national newspapers and information provided by public organizations. This group considered it significantly less important that information is provided from official sources, but instead considered it significantly more important that all opinions can be expressed, that the information source is independent from state, politics and economy and to know the author of the information personally. This could indicate a general skepticism about official sources. Similar, although not directly comparable, patterns were shown by Ali et al. (2020) among US adults during the Covid-19 pandemic, where trust in government websites and certain mainstream media was correlated with better knowledge about Covid-19.

As information ecology is constantly changing, the Covid-19 pandemic is the first global pandemic to take place in the current information environment, with all the new possibilities of the Internet, such as easier access to global information and user-created content in social media. The results confirm that traditional media channels continue to play an important role in the situation of a crisis, where the increased demand for reliable information in the German-speaking countries runs alongside increased use of reliable information sources like public broadcasters. This underpins the importance of these institutions, particularly in the situation of a crisis, also in context of ongoing critical discussion in the German-speaking countries about the need of public broadcasting (Lucht, 2006). However, the findings also show that increased frequency in media channel use happened mainly among irregular users, while non-users continued to stay away, which implies that also other communications channels like social media can be important official communication channels in the situation of a public health crisis. That the participants reported to also use social media intentionally to follow official and quality media sources shows that social media use as information source is not essentially associated with lower information quality and reliability, as concerned by Schäfer et al. (2020) regarding the high priority of social media content as information source for German university students in the Covid-19 crisis. That participants with awareness of health risks in terms of belonging to a risk group of Covid-19 showed to have a higher perception of the importance of health-related knowledge acquisition, implies that clear communication of the individual risks associated with contracting Covid-19 might be key for the success of related communication campaigns, e.g. on Covid-19 vaccinations.

This work comes with several limitations, which in turn provide avenues for future research: First, the sample size of the study of only 308 participants from the German-speaking countries is not representative for the whole population in these countries. The sample includes an especially high number of participants with a university degree. However, as the results resemble usage patterns from previous studies based on representative samples (Frees *et al.*, 2019; Breunig and Hottmannspötter, 2019), specifically the more frequent use of newspapers and television and less frequent use of streaming services and social media by older users, it can be assumed that the results bear some generalizability. Second, the applied survey could only find the subjective perception of the participants, which might not fully reflect their actual behavior. Third, as the questionnaire was online for several weeks, there might also have been changes during the duration of the survey due to the high dynamic of the crisis, for example the main topics of interest for seeking Covid-19-related information might have shifted during the survey period.  Fourth, as the answers to the open questions have shown, some participants consume media outlets such as newspapers or television solely by following their social media channels. While the questionnaire used in this study did distinguish between the most common social media services in the German-speaking countries, it did not ask in detail which content is consumed through social media. This shows the need to use more differentiated questions regarding social media use in future studies.

      Due to the global scale of the Covid-19 pandemic, future research might be extended to further countries and languages. As the Covid-19 crisis continues, it will also be of interest to see how the crisis impacts information behavior in the long term.

Table I.

*Composition of the Sample*

| Demographic variable | Categories | N | % |
|---|---|---|---|
| Gender | Female | 184 | 59.74% |
| | Male | 119 | 38.64% |
| | Diverse | 2 | 0.65% |
| | No answer | 3 | 0.97% |
| | Total | 308 | 100% |
| Age | <18 | 1 | 0.32% |
| | 18–29 | 91 | 29.55% |
| | 30–39 | 69 | 22.40% |
| | 40–49 | 58 | 18.83% |
| | 50–59 | 59 | 19.16% |
| | >60 | 29 | 9.42% |
| | No answer | 1 | 0.32% |
| | Total | 308 | 100% |
| Federal State | Baden-Wuerttemberg | 20 | 6.49% |
| | Bavaria | 41 | 13.31% |
| | Berlin | 27 | 8.77% |
| | Lower Saxony | 106 | 34.42% |
| | North Rhine-Westphalia | 40 | 12.99% |
| | Other/Not Germany | 73 | 23.70% |
| | No answer | 1 | 0.32% |
| | Total | 308 | 100% |
| Graduation | Qualified secondary school certificate | 13 | 4.22% |
| | Secondary school leaving certificate | 15 | 4.87% |
| | Abitur/A-Level | 50 | 16.23% |
| | University degree | 216 | 70.13% |
| | Other | 12 | 3.90% |
| | No answer | 2 | 0.65% |
| | Total | 308 | 100% |
| Working from Home | Yes | 222 | 72.08% |
| | No | 74 | 24.03% |
| | No answer | 12 | 3.90% |
| | Total | 308 | 100% |
| Belonging to High-Risk Group | Yes | 61 | 19.81% |
| | No | 235 | 76.30% |
| | No answer | 12 | 3.90% |
| | Total | 308 | 100% |

Table II.

*Comparison of Media Usage Before and During the Covid-19 Crisis*

| | Pre-Covid Mean | Pre-Covid SD | Covid Mean | Covid SD | Mean Difference | Z | P< | r |
|---|---|---|---|---|---|---|---|---|
| Local newspaper (offline & online) | 3.35 | 2.106 | 3.06 | 2.136 | 0.286 | -4.378 | < .001*** | .25 |
| National newspaper (offline & online) | 2.92 | 2.014 | 2.51 | 1.904 | 0.406 | -5.948 | < .001*** | .34 |
| Private television | 5.30 | 2.091 | 5.13 | 2.232 | 0.169 | -2.996 | .003** | .17 |
| Public television | 2.69 | 1.858 | 2.24 | 1.789 | 0.445 | -6.510 | < .001*** | .37 |
| Public organizations | 5.56 | 1.727 | 3.02 | 1.744 | 2.542 | -13.565 | < .001*** | .77 |
| International sources (radio, broadcast, newspaper) | 3.76 | 2.131 | 3.24 | 2.085 | 0.519 | -6.164 | < .001*** | .35 |
| YouTube | 4.55 | 2.221 | 4.42 | 2.289 | 0.127 | -1.598 | .110 | .09 |
| Facebook | 4.81 | 2.514 | 4.85 | 2.515 | -0.042 | -0.602 | .547 | .03 |
| Twitter | 5.07 | 2.555 | 4.93 | 2.658 | 0.140 | -2.891 | .004** | .16 |
| Instagram | 5.12 | 2.569 | 5.08 | 2.568 | 0.036 | -0.714 | .475 | .04 |
| Music streaming, podcasts | 4.83 | 2.433 | 4.68 | 2.490 | 0.159 | -2.413 | .016* | .14 |
| Physical meetings and discussions | 2.04 | 1.520 | 3.74 | 2.110 | -1.701 | -11.474 | < .001*** | .65 |
| Online communication with acquaintances and friends | 2.63 | 1.746 | 2.16 | 1.450 | 0.471 | -5.619 | < .001*** | .32 |

*Note.* 1: daily; 7: never; * $p < .05$, ** $p < .01$; *** $p < .001$; $r >= .1$ small effect, $r >= .25$ medium effect, $r >= .4$ strong effect.

Table III.

*Pearson Correlations of Reported Media Use and Usage Criteria with Age, Education and Satisfaction with Information Supply during the Covid-19 Crisis*

| Media Channels (Scale: 1 – daily / 7 – never) | Age | Education | Satisfaction |
|---|---|---|---|
| Local newspaper (offline & online) | -.187** | .072 | -.008 |
| National newspaper (offline & online) | -.112* | -.074 | .208*** |
| Private television | -.010 | .036 | -.081 |
| Public television | -.170** | -.032 | .228*** |
| Public organizations | -.037 | -.075 | .169** |
| International sources (radio, broadcast, newspaper) | -.105 | -.006 | .076 |
| YouTube | .197** | .011 | -.074 |
| Facebook | .160** | -.035 | -.048 |
| Twitter | .111 | -.009 | .157** |
| Instagram | .324*** | .019 | .020 |
| Music streaming, podcasts | .247*** | -.080 | .066 |
| Physical meetings and discussions | -.016 | .008 | -.057 |
| Online communication with acquaintances and friends | .012 | .113* | -.018 |
| *Criteria for choosing information sources (Scale: 1 – Applies to a great extent / 7 – Does not apply at all)* | Age | Education | Satisfaction |
| Availability of topics from my region | -.048 | -.030 | -.120* |
| Journalistic quality | -.042 | .085 | .072 |
| Credible information | -.029 | .030 | .100 |
| Interesting facts from research, history | -.078 | .045 | .089 |
| Background information on many topics | -.123* | .062 | -.015 |
| Communication of social values | -.010 | .026 | .060 |
| All opinions can be expressed | -.026 | .075 | -.263*** |
| Reliable help in daily life | .012 | .093 | -.032 |
| Support for the formation of political opinion | -.010 | .011 | .098 |
| Independence from state, politics and economy | -.144* | .173** | -.223*** |
| Interesting for the whole family | -.036 | .234*** | -.121* |
| Good entertainment | .100 | .076 | -.030 |
| Current live information | .197** | -.036 | .072 |
| Little advertising | -.066 | .121* | .018 |
| Knowing the author personally | -.156** | .158** | -.160** |
| Personal recommendation | -.053 | .128* | -.059 |
| Information from official sources | .161** | .006 | .178** |

*Note:* * *p* < .05; ** *p* < .01; *** *p* < .001.

Figure 1.

*Media Usage Distribution in Percent*

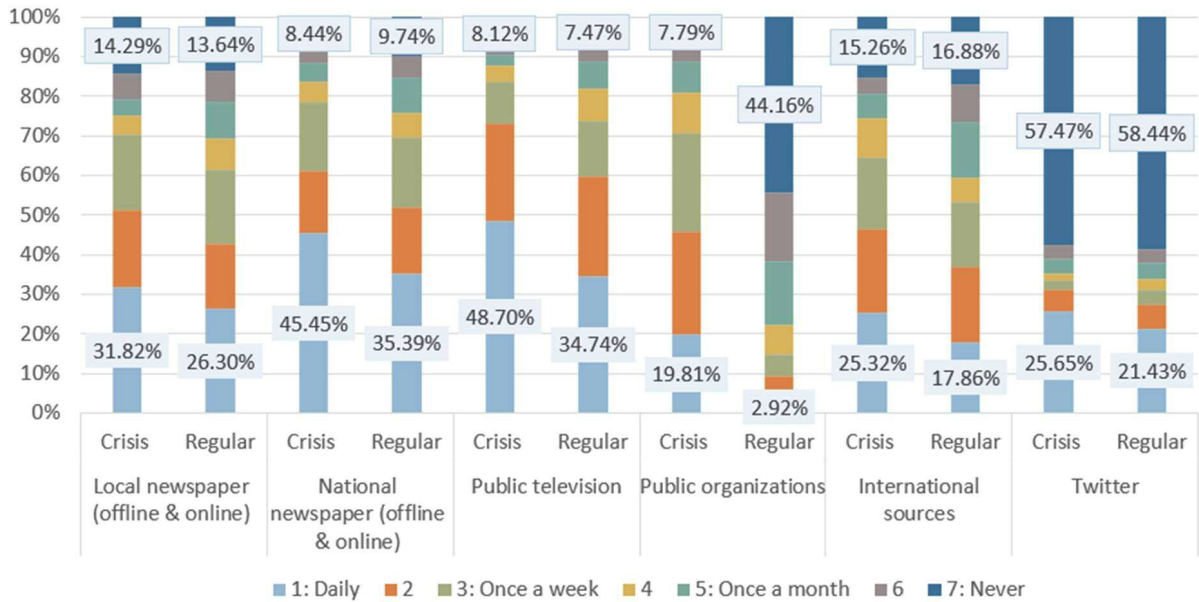

Figure 2.

*Criteria for Choosing Information Sources During the Covid-19 Crisis*

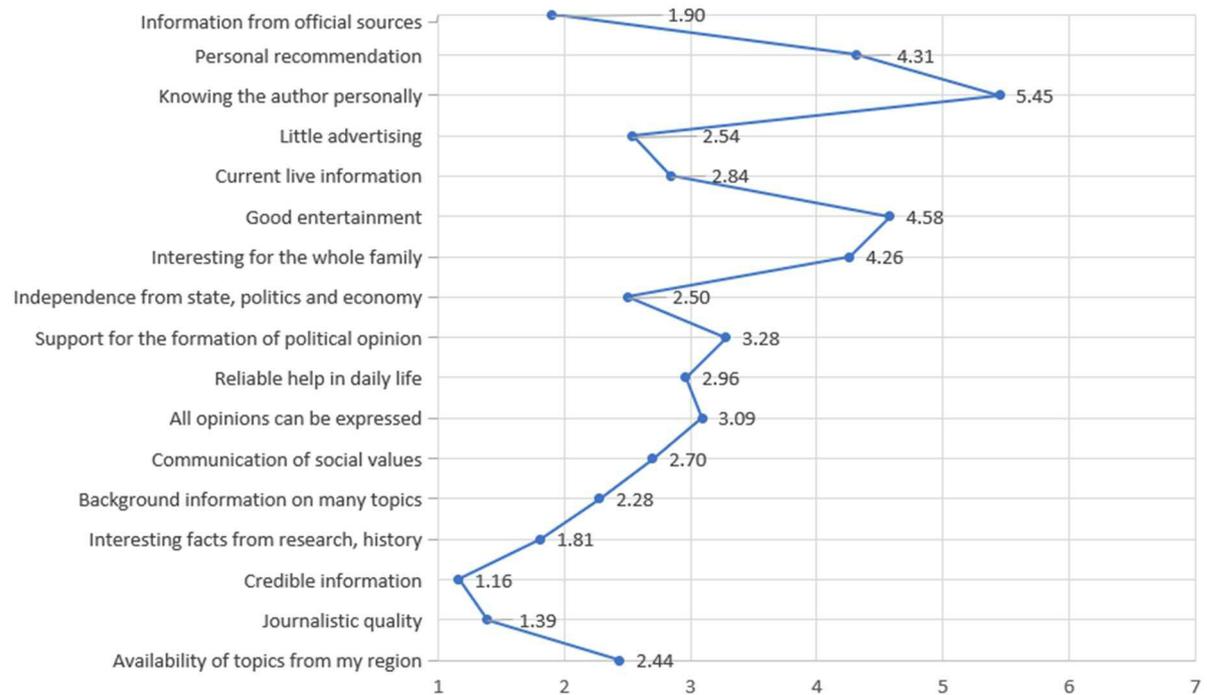

*Note:* Level of agreement (1: Applies to a great extent; 4: Neutral; 7: Does not apply at all).

Figure 3.

*Reasons for Information Seeking During the Covid-19 Crisis*

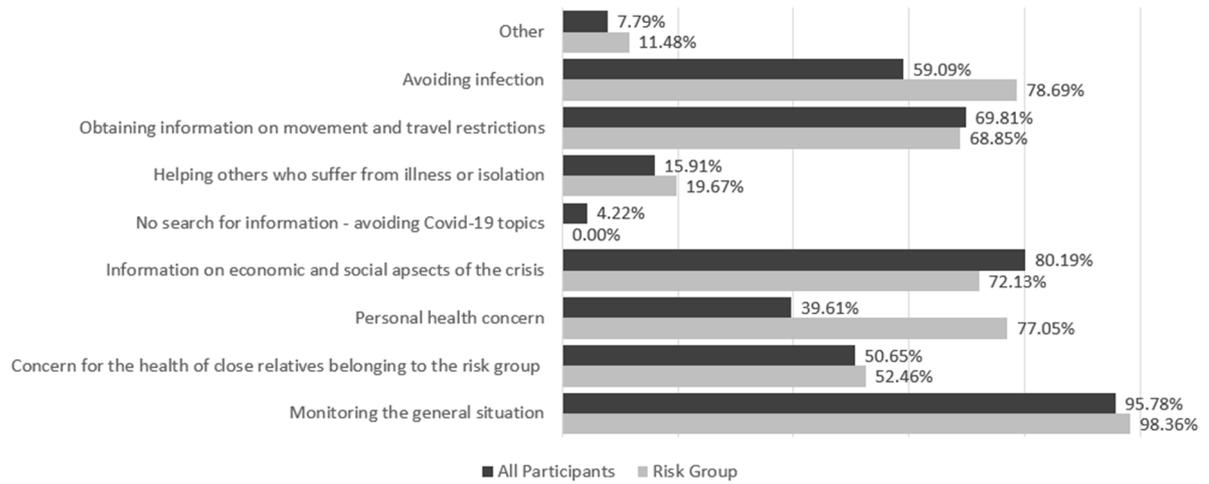